\documentclass[aps,pra,floatfix,superscriptaddress,twocolumn]{revtex4-2}

\usepackage{amsmath,amssymb,graphicx,color,braket,xcolor}
\usepackage[ocgcolorlinks,colorlinks=true,linkcolor=blue,citecolor=red,linktocpage=true]{hyperref}
\usepackage{bbm}

\usepackage{physics}

\begin{document}

\title{Thermodynamics of hybrid quantum rotor devices}
\author{Heather Leitch}
\author{Kenza Hammam}
\author{Gabriele De Chiara}
\affiliation{Centre for Quantum Materials and Technology, School of Mathematics and Physics,
Queen’s University Belfast, Belfast BT7 1NN, United Kingdom}

\date{\today}
\begin{abstract}
    We investigate the thermodynamics of a hybrid quantum device consisting of two qubits collectively interacting with a quantum rotor and coupled dissipatively to two equilibrium reservoirs at different temperatures. By modelling the dynamics and the resulting steady state of the system using a collision model, we identify the functioning of the device as a thermal engine, a refrigerator or an accelerator. In addition, we also look into the device's capacity to operate as a heat rectifier, and optimise both the rectification coefficient and the heat flow simultaneously. Drawing an analogy to heat rectification and since we are interested in the conversion of energy into the rotor's kinetic energy, we introduce the concept of angular momentum rectification which may be employed to control work extraction through an external load.    
\end{abstract}

\maketitle

\section{Introduction}
The exploration of quantum thermal machines is crucial in comprehending the intricacies of energy exchange on a quantum level, as evidenced by extensive research~\cite{Kosloff2014,binder2018thermodynamics,MitchisonCP2019,MyersAVS2022}. In particular, one would like to determine the role of genuine quantum effects, such as quantum coherence and entanglement, on the laws of thermodynamics \cite{sai2016,binder2018thermodynamics}. One of the main features of quantum thermal machines is their ability to outperform their classical counterparts \cite{LatuneSR2019,Hammam_2022,wang2019finite}, which can lead to a new generation of highly performing and energy-efficient thermal devices for novel quantum technologies applications.

Autonomous thermal machines \cite{TonnerPRE2005,Popescu2010,SilvaPRE2016, Mitchison_2016,Hammam_2021} offer a remarkable opportunity to study these quantum effects since they are known for featuring the lowest level of control and energetic cost \cite{Clivaz2019}, unlike reciprocating engines that operate in a cycle which consists of discrete strokes, for example, the four stroke engines that are employed in the Otto and Carnot cycles \cite{kosloffPRX2015}. 

There are several models of autonomous thermal machines that utilise various quantum systems. 
 For instance, a recent implementation of an autonomous quantum absorption refrigerator involved trapped ions~\cite{maslennikov2019quantum}, and proposals for implementation exist in other platforms like circuit-QED architecture~\cite{QED2016} and quantum dots \cite{Fazio2013,BohrBrask2022operational}.

One of the main open problems in the area of quantum thermal machines is how to convert the work generated by the machine into mechanical motion. One promising avenue consists of attaching a quantum rotor to the engine. The work thus converted into motion, can then be extracted by means of a dissipative load \cite{RouletPRE2017,seah2018work,seah2018quantum,RouletQST2018,PueblaPRR2022} or the rotor may act as a quantum flywheel, storing energy for later use~\cite{LevyPRA2016,vonLindenfelsPRL2019,VanHorneNPJ2020,CulhanePRE2022,martins2023rydberg}.

\begin{figure}[t]
\includegraphics[width=0.9\columnwidth]{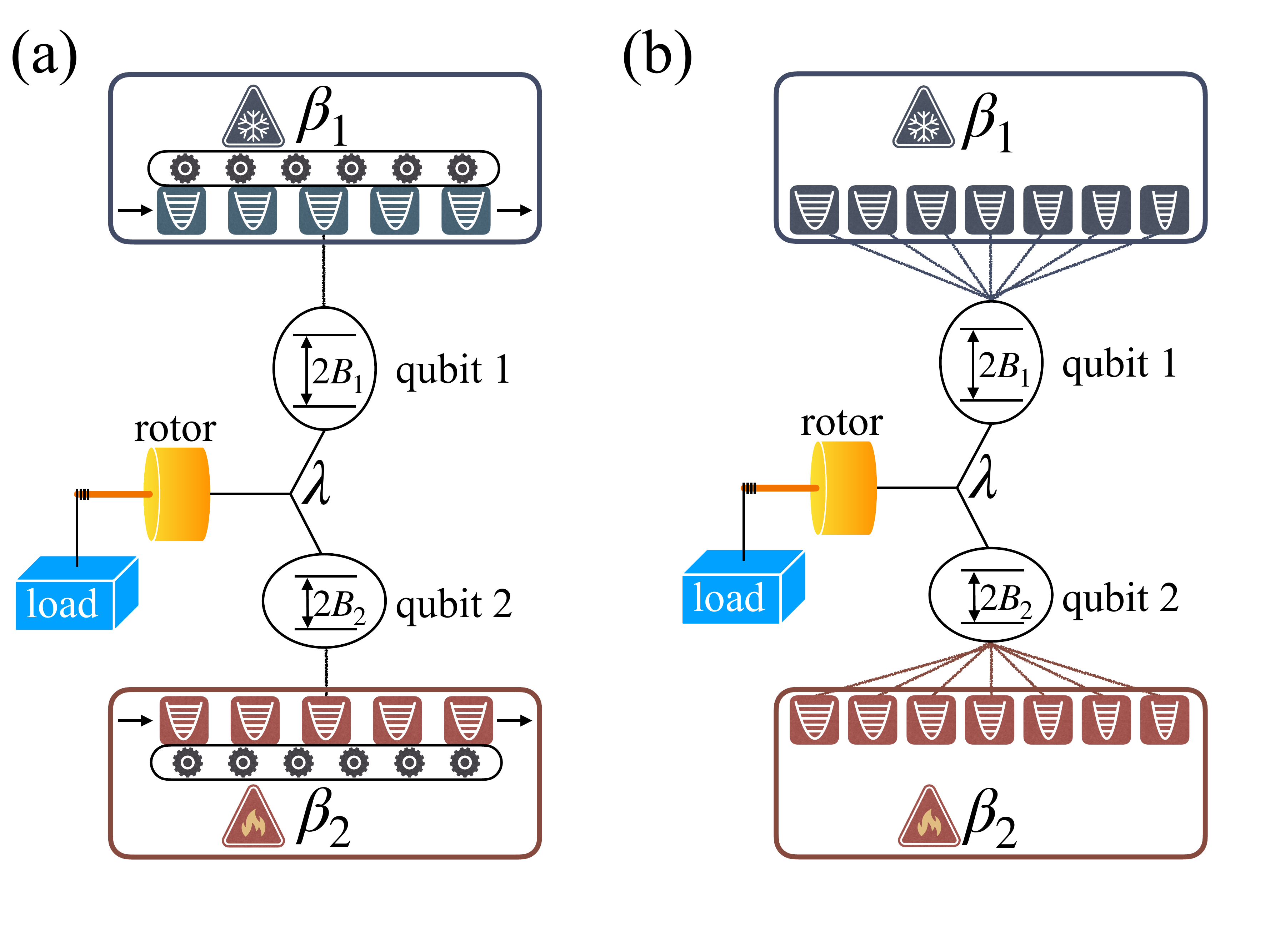}
\caption{Schematic representation of the two qubit-one rotor system. The three subsystems are all coupled to each other with a coupling strength $\lambda$ as described by Eq. \eqref{eq3wayinteraction}. (a) Collision model: a series of quantum harmonic oscillators initially in thermal equilibrium at inverse temperature $\beta_1$ ($\beta_2$) interact sequentially with qubit 1 (2), $\beta_2<\beta_1$.
(b) Spin-boson model: Qubit 1 (2) is permanently coupled to an equilibrium bath at inverse temperature $\beta_1$ ($\beta_2$).}
\label{fig:diagram}
\end{figure}

In this paper, we look at a hybrid system consisting of two qubits and a rotor, see Fig.~\ref{fig:diagram}. The system is coupled to two equilibrium baths at different temperatures through each qubit. We assume a dissipative load is coupled to the rotor, enabling us to extract work from the setup as well as bring the system to a steady state in which the rotor acquires a terminal angular momentum. 

In the first version of the model, Fig.~\ref{fig:diagram}(a), the environments are modelled using the so called collision models or repeated interactions models \cite{CampbellEPL2021,CiccarelloPR2022,Cusumano2022} which have recently been  implemented experimentally~\cite{JinPRA2015,CuevasSR2019,GarciaPerez2020,MeloPRA2022,cech2023thermodynamics}.
In the limit of a short collision time, the evolution of the system is described by a local master equation (LME) whose thermodynamics is well understood and does not lead to any inconsistency~\cite{DeChiaraNJP2018}. 
 Within this framework, we find the conditions required for the device to operate as an engine, refrigerator and accelerator.  

In the second version of the model, we also investigate the device's potential to act as a heat rectifier, which would result in a non-symmetrical heat flow.
To this end, see Fig.~\ref{fig:diagram}(b), the environment, made of an infinite number of quantum harmonic oscillators, is permanently coupled to the system. In this framework, the system dynamics is well described by a global master equation (GME)~\cite{Gonzalez2017,HoferNJP2017}.  We evaluate and optimise the performance of the device in this capacity together with the maximum amount of heat flow. Note that, in contrast to Ref.~\cite{RouletPRE2017}, the presence of two qubits with different energy separations is crucial to achieve heat rectification.

Additionally, motivated by the conversion of heat into mechanical motion, we introduce the concept of angular momentum rectification. This phenomenon is observed when swapping the temperatures of the external baths, resulting in a significant change in the rotor's motional response. Similar to heat rectification, we evaluate this feature and demonstrate the device's employability as a switch when coupling a load to the rotor, allowing for control of its motion.

Overall, our findings showcase the versatility of this hybrid quantum device and its potential to perform a range of useful functions, which could have significant implications for the experimental development of advanced quantum technologies, see for instance Refs.~\cite{XiangRMP2013,Kurizki2015,Clerk2020}.

In Sec.~\ref{sec: model}, we start with a description of the proposed model before deriving the corresponding  master equations to describe the open dynamics of the system.  We discuss several measures of work for the rotor in Sec.~\ref{sec:work} and possible thermodynamic operations that the system may perform in Sec.~\ref{sec:operations}. The definitions of the rectification parameter and rectification performance measure for both heat flow and angular momentum are given in Sec.~\ref{sec:rectification}. The results of our calculations are displayed in Sec.~\ref{sec:results} and we conclude in Sec.~\ref{sec:conclusions}.

\section{Model}\label{sec: model}
Our system consists of two qubits interacting with a rotor such that the system's Hamiltonian is ($\hbar=1$):
\begin{equation}
H_S = H_{S,0}+ H_{S,I},
\end{equation}
where the first term is composed of the free Hamiltonian of the two qubits and the kinetic energy of the rotor
\begin{equation}
H_{S,0} = B_1 \sigma_1^z + B_2 \sigma_2^z + \frac{L_z^2}{2I},
\end{equation}
with $L_z$ and $I$ the rotor's angular momentum projection along $z$ and moment of inertia, respectively. The operators $\sigma_i^x,\sigma_i^y, \sigma_i^z$ are the Pauli operators for the qubits while $B_i$ are the applied magnetic fields, with $i=1,2$. The collective three-way interaction depicted in Fig.~\ref{fig:diagram}  is described by the so-called ``quantum mill" Hamiltonian \cite{gilz2013generalized,seah2018work}
\begin{equation}\label{eq3wayinteraction}
H_{S,I}  = \lambda (\sigma_1^+\sigma_2^- e^{i\varphi} + \sigma_1^-\sigma_2^+ e^{-i\varphi}) ,
\end{equation}
where $\sigma_i^{\pm}=(\sigma_i^x+i\sigma_i^y)/2$ are ladder operators and $\varphi$ is the angle operator canonically conjugated to $L_z$, i.e. $[e^{i \varphi},L_z]=-e^{i \varphi}$~\cite{RouletPRE2017}. The three-body Hamiltonian $H_{S,I}$ allows energy transfer from one qubit to another concurrently with the rotation of the rotor.
Three-body interactions have been proposed and experimentally realised in different quantum platforms, including ultracold atoms, trapped ions, Rydberg ions and superconducting circuits~\cite{PachosPRL2004,AbdiPRL2015,FengPRA2020,Liu2020,martins2023rydberg,MenkePRL2022}. Though three-body interactions are not necessary to realise thermal machines~\cite{HewgillPRE2020}, they often lead to more powerful and efficient machines~\cite{Popescu2010}.

Qubit 1 is coupled to a cold bath at inverse temperature $\beta_1$ and qubit 2 is coupled to a hot bath at inverse temperature $\beta_2$, with $\beta_2<\beta_1$. The rotor is coupled to a dissipative load that allows it to reach a stationary rotation rate without accelerating indefinitely.
In the weak system-bath coupling regime, we assume the reduced dynamics of the system to be governed by a Lindblad-Markovian master equation:
\begin{equation}
\dot{\rho}_S =-i \comm{H_S}{\rho_S}+\mathcal{L}_1[\rho_S] + \mathcal{L}_2[\rho_S]+ \mathcal{L}_r[\rho_S],
\end{equation}
where $\mathcal{L}_i$, $i=1,2,r$, are the Lindblad superoperators associated with the interaction of the  system with the cold bath, the hot bath, and the dissipative load of the rotor, respectively.

In the rest of the paper, we consider two models (see Fig.~\ref{fig:diagram}). In the first model (collision model, Fig.~\ref{fig:diagram}(a)), the environment is modelled as a series of quantum harmonic oscillators sequentially interacting with the system. The resulting dynamics in the limit of short collisions lead to an LME. In the second model (spin-boson model, Fig.~\ref{fig:diagram}(b)), each qubit is permanently coupled to an infinite ensemble of quantum harmonic oscillators in thermal equilibrium. The resulting dynamics leads to a GME. 

\subsection{Collision model}\label{sec:local ME}
In this section we introduce the first model, depicted in Fig.~\ref{fig:diagram}(a), in which each qubit is coupled to its environment through repeated interactions.
Each bath is modelled by an ensemble of identical bosonic modes such that their Hamiltonian is $H_B=H_{B,1}+H_{B,2}$, where
\begin{equation}
H_{B,i} = 2B_i \sum_{k=1}^{\infty} a_{i,k}^{\dagger}a_{i,k}.
\end{equation} 
 The index $k$ labels the bath’s field modes with $a_{i,k} (a_{i,k}^{\dagger})$ being their annihilation (creation) operators, fulfilling the usual bosonic commutation relations: $[a_{i,k}, a_{i',k'}^{\dagger}]=\delta_{ii'}\delta_{kk'}$ and $[a_{i,k}, a_{i',k'}]=0$. Each quantum harmonic oscillator interacts with the corresponding qubit for a small time $\tau$ and then it is discarded.
Each oscillator in the cold [hot] bath is coupled to qubit 1 [2] with a coupling strength $g_1$ [$g_2$] so that the system-bath interaction Hamiltonian is given by
\begin{eqnarray}
H_{I} &=&\sum_{i=1,2}\sum_{k=1}^{\infty} g_{i}\left[ \sigma_{i}^- a_{i,k}^{\dagger} + \sigma_{i}^+ a_{i,k} \right].
\end{eqnarray}

Following the collision model approach, as in \cite{DeChiaraNJP2018,CampbellEPL2021,CiccarelloPR2022}, we define
\begin{equation}
H = H_S +H_B + \frac{1}{\sqrt{\tau}}H_I .
\end{equation}
We prepare the baths in thermal states at inverse temperature $\beta_{i}$ ($i=1, 2$) 
\begin{equation} 
\rho_{B,i} =  \frac{e^{-\beta_{i}H_{B,i}}}{\text{Tr}[e^{-\beta_{i}H_{B,i}}]},
\end{equation}
and we define $\rho_B=\rho_{B,1}\otimes \rho_{B,2}$.

A collision occurring at a time $t$ can be described by a unitary operator $\mathcal{U} = e^{-i H\tau}$ such that the state of the system after the collision is expressed as
\begin{equation} \label{eq:rhocol}
\rho_S(t+\tau) = \text{Tr}_B \left[ \mathcal{U}\;\rho_S(t)\otimes \rho_B\;\mathcal{U}^{\dagger} \right],
\end{equation}
where $\rho_S(t)$ is the state of the system before the interaction with the thermal baths.
Then, we apply the Baker-Campbell-Haussdorff formula
\begin{eqnarray}\label{eq BH relation}
e^{iH\tau}Ae^{-iH\tau} &=& A +i\tau\comm{H}{A}+\frac{(i\tau)^2}{2!}\comm{H}{\comm{H}{A}} \nonumber\\
&+& \dots +\frac{(i\tau)^n}{n!}\comm{H}{\dots\comm{H}{\comm{H}{A}}}+\dots , \nonumber\\
\end{eqnarray}
which leads to the expansion of Eq. \eqref{eq:rhocol} in $\tau$
\begin{eqnarray}
\rho_S(t+\tau) &= & \rho_S(t) -i\tau \text{Tr}_B  \comm{H}{\rho_S(t)\otimes \rho_B}  \nonumber\\
& -& \frac{\tau}{2} \text{Tr}_B  \comm{H_I}{\comm{H_I}{\rho_S(t)\otimes\rho_B}}  + O(\tau^2).\nonumber\\
\end{eqnarray}
By dividing both sides by $\tau$ and taking the limit of $\tau\to 0$, we find the time derivative 
\begin{equation}
\dot{\rho}_S = \lim_{\tau \rightarrow 0} \frac{\rho_S(t+\tau)-\rho_S(t)}{\tau},
\end{equation}
which fulfils the LME (dropping the explicit time dependence of $\rho_S$)
\begin{eqnarray}\label{eq:ME}
\dot{\rho}_S=-i \comm{H_S}{\rho_S}+\sum_{i=1,2,r}\mathcal{L}_i^{\text{(loc)}}[\rho_S].
\end{eqnarray}

The local thermal Lindblad superoperators acting on qubits $i=1,2$ are given by:
\begin{equation}
\mathcal{L}_i^{\text{(loc)}}[\rho_S]=g_i^2 \left(n_i\mathcal{D}[\sigma_i^+]+ (n_i+1)\mathcal{D}[\sigma_i^-]\right)
\end{equation}
where $\mathcal{D}[O] = O \rho_S O^{\dagger} - \frac{1}{2}\{O^{\dagger}O,\rho_S\}$ represent the usual Lindblad dissipators \cite{breuer2002theory}.
For later convenience, we define the coupling constants between the system and each bath as
\begin{eqnarray}
g_1 &=& g(1 - \chi), \nonumber\\
g_2 &=& g(1 + \chi),
\end{eqnarray}
so that varying $-1 < \chi < 1$ allows us to study the effect of the difference in coupling strength of the cold and hot baths. 
The thermal occupations $n_i$ are related to the baths' inverse temperatures by:
\begin{equation}
n_i = [e^{2\beta_i B_i}-1]^{-1}.
\end{equation}

Employing a similar treatment for the rotor, the action of the dissipative load corresponds to the Lindblad term $\mathcal{L}_r$\cite{seah2018work,stickler2018rotational}
\begin{eqnarray}
\label{eq:dissload}
\mathcal{L}_r^{\text{(loc)}}[\rho_S] &=& \frac{2I\gamma}{\beta_r}
\left\{ 
\mathcal{D}\left[ \cos\varphi - \frac{i\beta_r \sin\varphi L_z}{4I} \right]
\right .
\\
&+& \left .
\mathcal{D}\left[ \sin\varphi + \frac{i\beta_r \cos\varphi L_z}{4I} \right] 
\right\},
\nonumber
\end{eqnarray}
where $\gamma$ is the dissipation rate and $\beta_r$ is the inverse temperature of the bath modelling the dissipative load. The effect of the dissipative load is twofold: it provides a frictional force allowing the rotor to reach a steady state, rather than continuing to accelerate, and it also allows the conversion of mechanical work from the rotor  into the load.
In the absence of coupling to the two qubits, the rotor coupled to the dissipative load will reach a thermal equilibrium state at inverse temperature $\beta_r$.

The thermodynamics of collision models can be derived from the corresponding microscopic model. 
Since the environmental harmonic oscillators are initially in thermal equilibrium, we obtain the heat flows as minus the energy change in the bath~\cite{DeChiaraNJP2018}:
\begin{eqnarray}
    \dot{Q}_{i}^{\text{(loc)}} =
    -\lim_{\tau\to 0} \frac 1\tau \text{Tr}\left [H_{B,i}\Delta\rho_{SB}\right ]
    = \text{Tr}[H_{S,0}\mathcal{L}_{i}^{\text{(loc)}}[\rho_S]],
    \nonumber
    \\
\label{eq:Qiapp}    
\end{eqnarray}
where $\Delta\rho_{SB}=\mathcal{U}\;\rho_S(t)\otimes \rho_B\;\mathcal{U}^{\dagger}-\rho_S(t)\otimes \rho_B$. After some algebra we obtain more explicitly
\begin{equation}
\dot{Q}_{i}^{\text{(loc)}} = 2B_{i} g_{i}^2 \left( n_{i}\expval{\sigma_{i}^- \sigma_{i}^{+}}_t - (n_i+1)\expval{\sigma_{i}^+ \sigma_{i}^{-}}_t\right),
\end{equation}
where
\begin{equation}
\expval{\sigma_{i}^+ \sigma_{i}^{-}}_t = \text{Tr}\left[\rho_S\sigma_{i}^+ \sigma_{i}^{-}  \right].
\end{equation}

Similarly, the dissipative load attached to the rotor absorbs the heat current
\begin{equation}\label{heatflowlocal}
\dot{Q}_{r}^{\text{(loc)}} =  \text{Tr}[H_{S,0}\mathcal{L}_{r}^{\text{(loc)}}[\rho_S]],    
\end{equation}
where $\mathcal{L}_{r}^{\text{(loc)}}$ is given in Eq.~\eqref{eq:dissload}.

For LMEs, besides the usual heat currents,  there may be an additional work source, due to the ``locality" of the collisions which model microscopically the dissipation, see Ref.~\cite{DeChiaraNJP2018}.
For the two qubits, this can be found as the total energy change of both the system and bath:
\begin{eqnarray}
\label{eq:WQ}
    \dot{W}^{(Q)} =\lim_{\tau\to 0} \frac 1\tau \text{Tr}\left [(H_S+H_{B})\Delta\rho_{SB} \right],
\end{eqnarray}
 which can be expressed as
\begin{eqnarray}
\dot{W}^{(Q)}&=&  -\frac 12 \lambda \left[g_1^2 (2n_1+1) + g_2^2 (2n_2+1) \right]\nonumber\\
&\times& \left[ \expval{\sigma_1^+ \sigma_2^- e^{i\varphi}}_t + \expval{\sigma_1^- \sigma_2^+ e^{-i\varphi}}_t \right].
\end{eqnarray}

The work injected by the local baths into the qubits can be also expressed as:
\begin{equation}\label{eqpowerlocal}
\dot{W}^{\text{(Q)}} = \text{Tr}\left[H_{S,I}\left(\mathcal{L}_1^{\text{(loc)}} [\rho_S] + \mathcal{L}_2^{\text{(loc)}}[\rho_S]\right)\right],
\end{equation}
while the corresponding extra work associated with the rotor's dissipation can be written as:
\begin{equation}\label{eqpowerrotor}
\dot{W}^{(\text{r})} = \text{Tr}\left[H_{S,I}\mathcal{L}_r^{\text{(loc)}} [\rho_S]\right].
\end{equation}

Using Eqs.~\eqref{eq:ME}, \eqref{eq:Qiapp}, \eqref{heatflowlocal}, \eqref{eq:WQ} and \eqref{eqpowerrotor}, it can be proven that the first law of thermodynamics holds \cite{DeChiaraNJP2018}:
 \begin{equation}
 \dot{U}_S = \dot{Q}_1^{\text{(loc)}}+\dot{Q}_2^{\text{(loc)}}+\dot{Q}_r^{\text{(loc)}} + \dot{W}^{\text{(Q)}} + \dot{W}^{\text{(r)}},
 \end{equation}
   where $\dot{U}_S={\rm Tr} \left[\dot \rho_S H_S \right ]$ is the rate of change of the system's energy (assuming $H_S$ to be time independent).
Since we are choosing $\gamma$ to be small, the magnitude of $\dot{W}^{\text{(r)}}$ is negligible compared to the heat flows and to $\dot{W}^{\text{(Q)}}$.

Throughout this paper, we are using the convention that heat flowing into the system or work done on the system, thus increasing its internal energy, has a positive sign and vice versa.

 \subsection{Spin-boson model}
 \label{sec:GME}
 In our second model, we consider a variant of the spin-boson model, in which each qubit is permanently coupled to an infinite set of quantum harmonic oscillators in thermal equilibrium, see Fig.~\ref{fig:diagram}(b). In the weak-coupling limit, the dynamics of the system can be described by 
 a global master equation with respect to the interaction between qubits. In this case, the jump operators are obtained by deriving the master equation in the basis of the system's full Hamiltonian's eigenstates. In order to derive the GME, we need to write the complete evolution of the system and the bath and then apply the Born-Markov and secular approximation in the eigenbasis of $H_S$. The rotor's dissipative load is, for simplicity, still described by its local form in Eq.~\eqref{eq:dissload} assuming the corresponding dissipation rate $\gamma$ is small compared to the coupling $\lambda$. For this reason the resulting equation, Eq.~\eqref{eq:globalme} should more accurately be dubbed a semiglobal master equation. A thorough derivation is reported in Appendix~\ref{app:globalME} and the resulting equation is:
 \begin{eqnarray}\label{eq:globalme}
\dot{\rho}_S &=& - i\comm{H_S}{\rho_S} + \mathcal{L}_1^{\text{(glob)}}[\rho_S] \nonumber\\ 
& & + \mathcal{L}_2^{\text{(glob)}}[\rho_S] + \mathcal{L}_r^{\text{(loc)}} [\rho_S].
\end{eqnarray}
where $\mathcal{L}_1^{\text{(glob)}}$, $\mathcal{L}_2^{\text{(glob)}}$ and $\mathcal{L}_r^{\text{(loc)}}$ are the superoperators  that account for the energy jumps induced by both the interaction and the thermal
baths.
 
 Under these assumptions, $\dot{W}^{\text{(Q)}}=0$.  Hence, the first law now reads
 \begin{equation}
 \dot{U}_S = \dot{Q}_1^{\text{(glob)}}+\dot{Q}_2^{\text{(glob)}}+\dot{Q}_r^{\text{(loc)}} + \dot{W}^{\text{(r)}}.
 \end{equation}
 The work correction $\dot{W}^{\text{(r)}}$, under the assumption of small $\gamma$, is much smaller than the magnitude of the heat flows, so we will neglect it in the following.

The heat flows in terms of the superoperators $\mathcal{L}_{i}^{\text{(glob)}}$ from the GME \eqref{eq:globalme}
\begin{equation}\label{heatflowglobal}
\dot{Q}_{i}^{\text{(glob)}} =  \text{Tr}\left[H_S\mathcal{L}^{\text{(glob)}}_{i}[\rho_S]\right],
\end{equation}
where $\mathcal{L}^{\text{(glob)}}_i$, $i=1,2$ now act on the combined Hilbert space of the two qubits and the rotor.

\section{Thermal machines}
In this section, we discuss the possible functionings of our setup as thermal devices. To do this we will be using the LME, the heat flows and power defined in Eqs. \eqref{heatflowlocal} and \eqref{eqpowerlocal}. First, we introduce in Sec.~\ref{sec:work} the rotor's mechanical work which is a measure of the amount of work that can be extracted from the rotor's kinetic energy. Then, in Sec.~\ref{sec:operations}, we list the possible modes in which the setup can operate as a thermal machine. Finally, we introduce heat rectification and angular momentum rectification as well as their measures in Sec.~\ref{sec:rectification}.

\subsection{The rotor's mechanical work}\label{sec:work}
When defining the work done by the rotor, there are several useful quantities that we may look at \cite{seah2018work,seah2018quantum}. 
The rate of change of the rotor's kinetic energy can simply be defined as:
\begin{eqnarray}
\label{eq:kineticpower}
\dot{W}_{\text{kin}} &=& \frac{d}{dt}\frac{\langle L_z^2 \rangle}{2I} = \frac{1}{2I}\text{Tr}[\dot{\rho}_S L_z^2]
\\
&=& \dot{W}_{\text{int}} +\dot{Q}_{BA}.
\end{eqnarray}
In the preceding expression, $\dot{W}_{\text{kin}}$ is split into two terms.
The first term represents the intrinsic power
\begin{equation}\label{eq intrinsic power}
\dot{W}_{\text{int}} = -\frac{i}{2I}\text{Tr}\left[ \comm{H_S}{\rho_S} L_z^2 \right]
\end{equation}
which is the rate of change in the kinetic energy that is due to the working medium exerting force on the rotor, and the second term is the back-action heat flow
\begin{equation}
\label{eq:QBA}
\dot{Q}_{BA} =   \frac{1}{2I} \text{Tr}\left[ (\mathcal{L}_1 [\rho_S] + \mathcal{L}_2[\rho_S] + \mathcal{L}_r [\rho_S]) L_z^2 \right],
\end{equation}
which is due to angular momentum diffusion in the rotor and does not contribute to useful work. In Eq.~\eqref{eq:QBA} the Lindblad superoperators may be from the local or global master equation.
Note that the kinetic energy could be disordered, as $\langle L_z^2 \rangle$ does not take into account the direction of the rotor's rotation. Also, the kinetic energy could come from the unwanted heating of the rotor. 

To resolve this issue, we can look instead at the net power, i.e. the rate of change in kinetic energy exclusively in either the clockwise or anticlockwise direction. This net power is defined as
\begin{equation}\label{eq net power}
\dot{W}_{\text{net}} = \frac{d}{dt}\frac{\langle L_z \rangle^2}{2I} = \frac{1}{I}\text{Tr}[\rho_S L_z]\text{Tr}[\dot{\rho}_SL_z].
\end{equation}
Note that at the steady state, when $\dot{\rho}_S=0$, the kinetic and net power are zero.

Another useful quantity to study the capacity of the system to do work is the quantum ergotropy, which quantifies the maximum amount of extractable work by means of unitary transformations \cite{allahverdyan2002mathematical,niedenzu2019concepts,touil2021ergotropy}:
\begin{equation}\label{eq:ergotropy}
\mathcal{W}_{\text{erg}} = \text{max}_{U} \left\{ \Tr [ H_S \left( \rho_S - U\rho_S U^{\dagger} \right) ]\right\}
\end{equation}
where $U$ is the unitary evolution of the system.

\subsection{Operations}\label{sec:operations}
The system may be able to operate in several different ways depending on the direction of energy flows. These different operations are summarised in Table \ref{table:1}. Note that for studying the possible operations of this system we will assume $\beta_1 > \beta_r > \beta_2$.
\begin{table}[t]
\centering
\begin{tabular}{|c | c | c | c|} 
\hline
Operation & Heat flow $\dot{Q}_1$ & Heat flow $\dot{Q}_2$ &Work $\dot{W}^{(Q)}$\\
 \hline
 Engine & $\dot{Q}_1 < 0$ & $\dot{Q}_2 > 0$ & $\dot{W}^{(Q)} < 0$  \\ [0.5ex] 
 \hline
 Refrigerator & $\dot{Q}_1 > 0$ & $\dot{Q}_2 < 0$ & $\dot{W}^{(Q)} > 0$ \\ [0.5ex]  
 \hline
 Accelerator & $\dot{Q}_1 < 0$ & $\dot{Q}_2 > 0$ & $\dot{W}^{(Q)} >0$\\ [0.5ex] 
 \hline
\end{tabular}
\caption{Possible achievable operations of the system and their conditions.}
\label{table:1}
\end{table}

To assess the performance of our setup when operating as an engine, we use the efficiency of the engine, defined as 
\begin{equation}
\eta= -\frac{\dot{W}^{(Q)}}{\dot{Q}_2}.
\end{equation}
Since we are using thermal baths, the efficiency is bounded by the Carnot value at the lowest and highest temperatures,
$\eta_C = 1- \beta_2/\beta_1$.

To assess the functioning of the system as a refrigerator we employ the coefficient of performance (COP), defined as \cite{Kosloff2014}
\begin{equation}
\text{COP} = \frac{\dot{Q}_1}{\dot{W}^{(Q)}},
\end{equation}
such that it is also upper-bounded by the Carnot COP, $COP_C = \beta_1/(\beta_1 - \beta_2)$.

\subsection{Rectification}\label{sec:rectification}
To study rectification, we will be using the GME and the heat flows that arise from it, as defined in Eq. \eqref{heatflowglobal}. The global approach of the system lends itself well to the study of heat rectification as we may calculate the global heat flows in and out of the system as there is no external work done on the system.

Heat rectification is a form of asymmetric transport through a device when the two heat baths are swapped. It occurs due to the nonsymmetric arrangement of the microscopic constituents of a system. In our case this may be due to the different magnetic fields $B_1$ and $B_2$ applied to the two qubits, as well as the coupling to the baths.
In order to study rectification \cite{khandelwal2023characterizing,zhang2009reversal,Bhandari_2021,riera2019dynamically,HewgillPRR2021}, we calculate the heat flow into one of the baths coupled to the qubits, e.g. $\dot{Q} = \dot{Q}_2$. We then swap the temperatures of the two heat baths, $\beta_1$ and $\beta_2$, keeping everything else the same (including the couplings to the baths) and recalculate the heat currents. The heat flow into the system is now
\begin{equation}
    \dot{Q}^{\text{(swap)}} = \dot{Q}_1^{\text{(swap)}}
\end{equation}
where (swap) indicates that the temperatures of the two heat baths have been swapped. We can then define the rectification parameter as
\begin{equation}\label{eq:heatR}
R= \left|\frac{\dot{Q} - \dot{Q}^{\text{(swap)}} }{\dot{Q} + \dot{Q}^{\text{(swap)}} }\right|  .
\end{equation}
If $R = 0$, this means that heat can flow equally in both directions and there is no rectification, $\dot{Q}^{\text{(swap)}} = \dot Q$. For $0 < R \leq 1$, the system acts as a rectifier. The maximum value $R = 1$ corresponds to an ideal rectifier where heat can only flow in one direction. 

The problem with this definition of rectification is that we may be able to achieve a value of $R$ close to $1$, but with heat flows that are almost zero, making the device not practically useful. To take into account both the rectification and the magnitude of the heat flow, we can introduce 
\begin{equation}\label{eq:heatJ}
J= \text{max}(|\dot{Q}|, |\dot{Q}^{\text{(swap)}}|)/(\lambda g^2)   
\end{equation}
which is the maximum of the two heat flows. Note that we have divided by $\lambda g^2$ so that $J$ is dimensionless. Then, we can define the rectification performance's measure as 
\cite{khandelwal2023characterizing}
\begin{equation}\label{eq:rec_cop}
\Gamma_{\alpha}  = \alpha R + (1-\alpha) J.
\end{equation}
The parameter $\alpha$ ($0 \leq \alpha \leq 1$) allows us to study the trade-off between rectification and heat flow.

\begin{figure}[t]
    \centering
    \includegraphics[width=0.95\columnwidth]{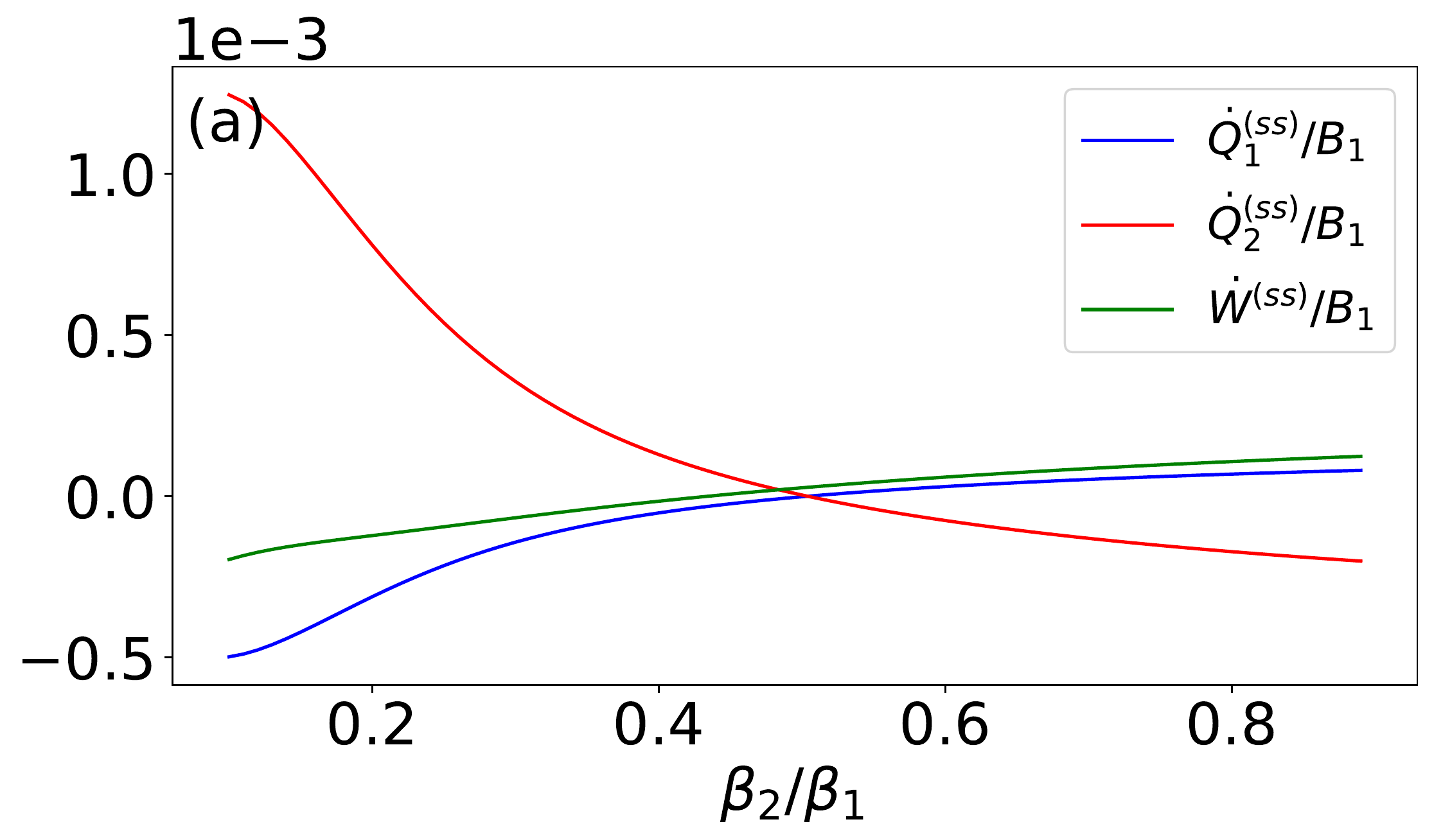}
    \\
    \includegraphics[width=0.48\columnwidth]{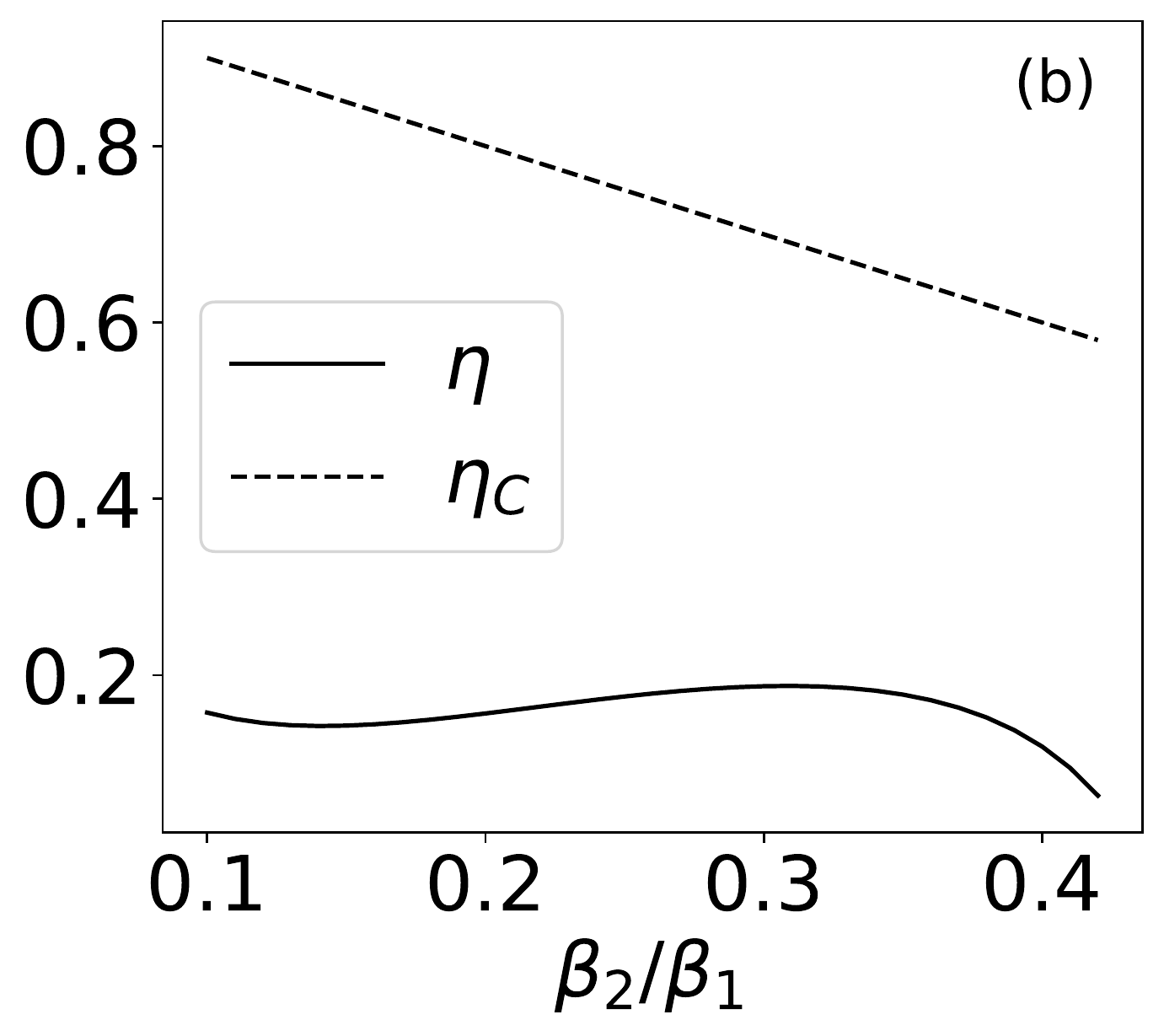}
 \includegraphics[width=0.45\columnwidth]{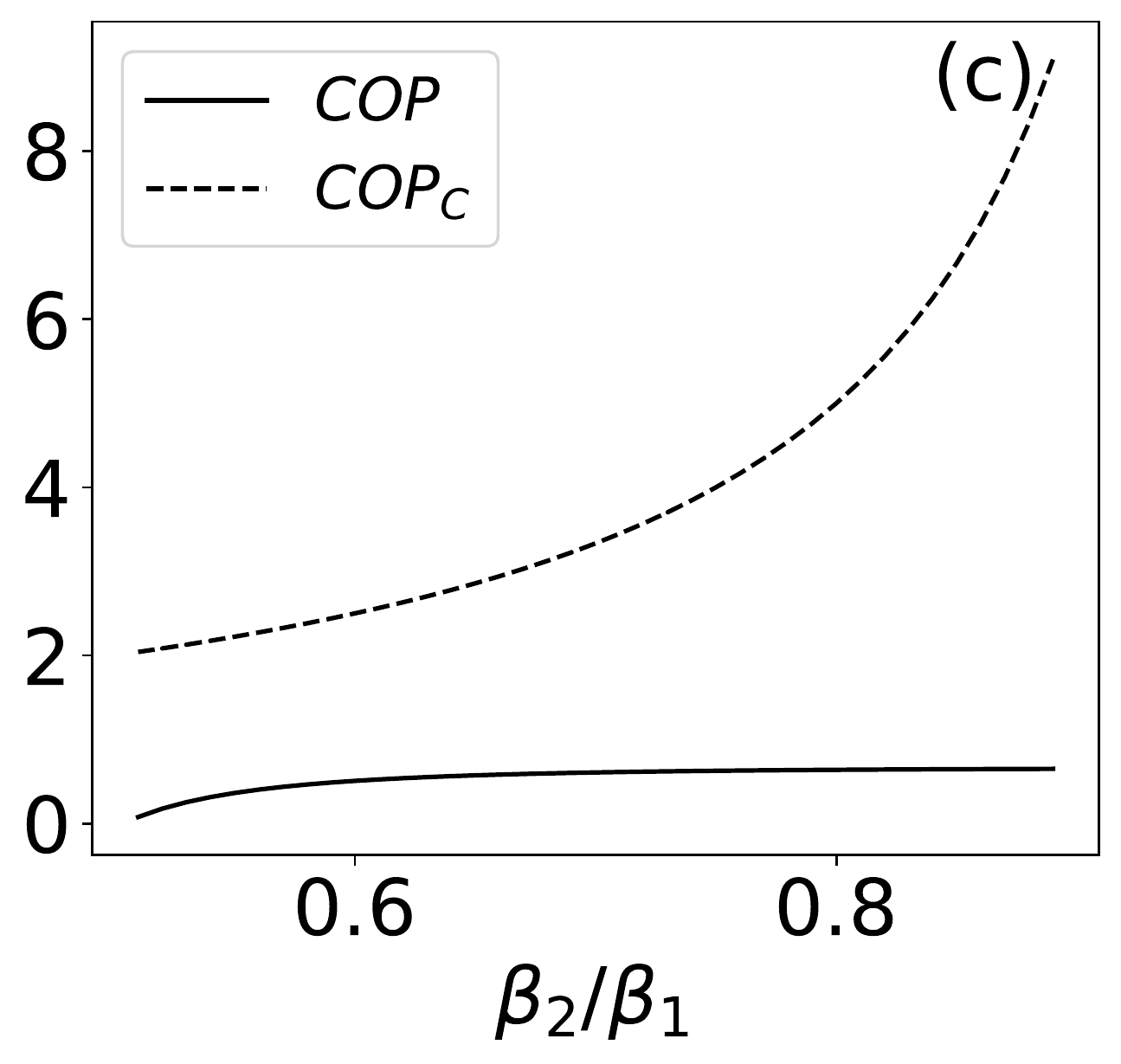}
    \caption{(a) Steady state heat flows and power  as a function of $\beta_2/\beta_1$. The device may operate as an engine, accelerator or refrigerator. 
    (b) Efficiency of the machine $\eta$ (solid line) while it operates as an engine and the corresponding Carnot efficiency $\eta_C$ (dashed line) as functions of $\beta_2/\beta_1$.
    (c) The COP (solid line) and the Carnot COP (dashed line) for the refrigerator regime as functions of $\beta_2/\beta_1$.
    The parameters are: $\beta_1=0.1, \beta_r=0.09, B_1=4, B_2=10, g=1, \chi=0, \lambda=0.1, \gamma=5\times10^{-5}$.}
    \label{fig:tri_operations}
\end{figure}

Not only can we study heat rectification, but we may also look at the rectification of the rotor's angular momentum. In fact, we can think of the rotor as a windmill that is being rotated by the heat flowing through the qubits and capable, due to its angular momentum, of lifting weights (modelled with the dissipative load). 

We define the angular momentum rectification coefficient to be of the form 
\begin{equation}\label{eq:angmomR}
R_{\rm angular} = \left|\frac{\expval{L_z} - \langle L_z^{\text{(swap)}}\rangle }{\expval{L_z} + \langle L_z^{\text{(swap)}}\rangle}\right|, 
\end{equation}
where $\expval{L_z}$ and $\langle L_z^{\text{(swap)}}\rangle$ are the steady-state expectation values of the 
angular momentum operator in one configuration and in the configuration obtained by swapping the temperatures.

As before, since we are interested not only in the angular momentum's rectification but also in its magnitude, we define
\begin{equation}
\label{eq:angmomJ}
J_{\rm angular}= \text{max}(|\expval{L_z}|, |\langle L_z^{\text{(swap)}}\rangle|).   
\end{equation}
We may then vary $0 \leq \chi \leq 1$ again and calculate the corresponding $R$ and $J$. The performance of the angular momentum rectification can be measured by the quantity
\begin{equation}
\label{eq:reccop}
\Gamma_{\alpha,\text{angular}}  = \alpha R_{\text{angular}} + (1-\alpha) J_{\text{angular}}.
\end{equation}

\section{Results}\label{sec:results}
In this section, we report and discuss the numerical results for the thermodynamics of our model. In order to simulate the rotor, we need to truncate its infinite Hilbert space. In Appendix \ref{app:truncation}, we discuss how we perform the truncation. All the obtained results bear no effect from the finiteness of the Hilbert space.

\subsection{Collision model: Thermal operations}\label{sec:3-way interaction}
Using the collision model, we calculate both the time-evolution of the system's density matrix and the steady state obtained by solving the equation $\dot\rho_S=0$. We verify that the long-time limit of the evolved density matrix agrees with the steady-state solution within the numerical accuracy.

In Fig.~\ref{fig:tri_operations}(a) we show the results for the steady state heat flows and power, by fixing the cold temperature $\beta_1$ and varying the hot temperature $\beta_2$, so that $0<\beta_2<\beta_r<\beta_1$ (see the figure's caption for all the other parameters). In Appendix \ref{sec:powerappendix} we investigate how the kinetic, intrinsic and net power vary over time.

We see that, depending on our choice of $\beta_2/\beta_1$, the system can operate in three different regimes: engine, accelerator or refrigerator.
From Fig.~\ref{fig:tri_operations}(a) we see that, for our choice of parameters, choosing $0 < \beta_2/\beta_1 < 0.43$ gives rise to an engine operation. The machine generates a work output ($\dot{W}^{(Q)}<0$) by using the natural heat flow from the hot bath ($\dot{Q}_2 >0$) while dumping a part of it into the cold bath ($\dot{Q}_1 <0$), see Table \ref{table:1}. 
In Fig.~\ref{fig:tri_operations}(b) we plot the corresponding steady-state efficiency of the engine. We remark that, while it is much lower than the Carnot efficiency, $\eta_C = 1 - \beta_2/\beta_1$, it reaches a maximum around $\beta_2/\beta_1 \simeq 0.31$ and gets close to $\eta_C$ at $\beta_2/\beta_1 \simeq 0.37$.

On the other hand, within the small range $0.43 < \beta_2/\beta_1 < 0.54$, the system acts as an accelerator which utilises a work input ($\dot{W}^{(Q)} > 0$) to speed up the transfer of heat from the hot bath ($\dot{Q}_2 > 0$) to the cold bath ($\dot{Q}_1 < 0$).

Finally, for $0.54 < \beta_2/\beta_1 < 0.9$, the machine functions as a refrigerator as work is consumed ($\dot{W}^{(Q)} > 0$) while heat from the cold bath ($\dot{Q}_1 > 0$) is transferred to the hot bath ($\dot{Q}_2 < 0$).
In Fig.~\ref{fig:tri_operations}(c), we can clearly see that the refrigerator's COP increases slowly with increasing $\beta_2/\beta_1$ and reaches its maximum at $\beta_2/\beta_1\simeq 0.54$.

Note that, as we have assumed $\beta_1>\beta_r>\beta_2$, it is not possible to choose $\beta_2/\beta_1>0.9$ for our choice of parameters.

\begin{figure}[t]
    \centering
    \includegraphics[width=0.95\columnwidth]{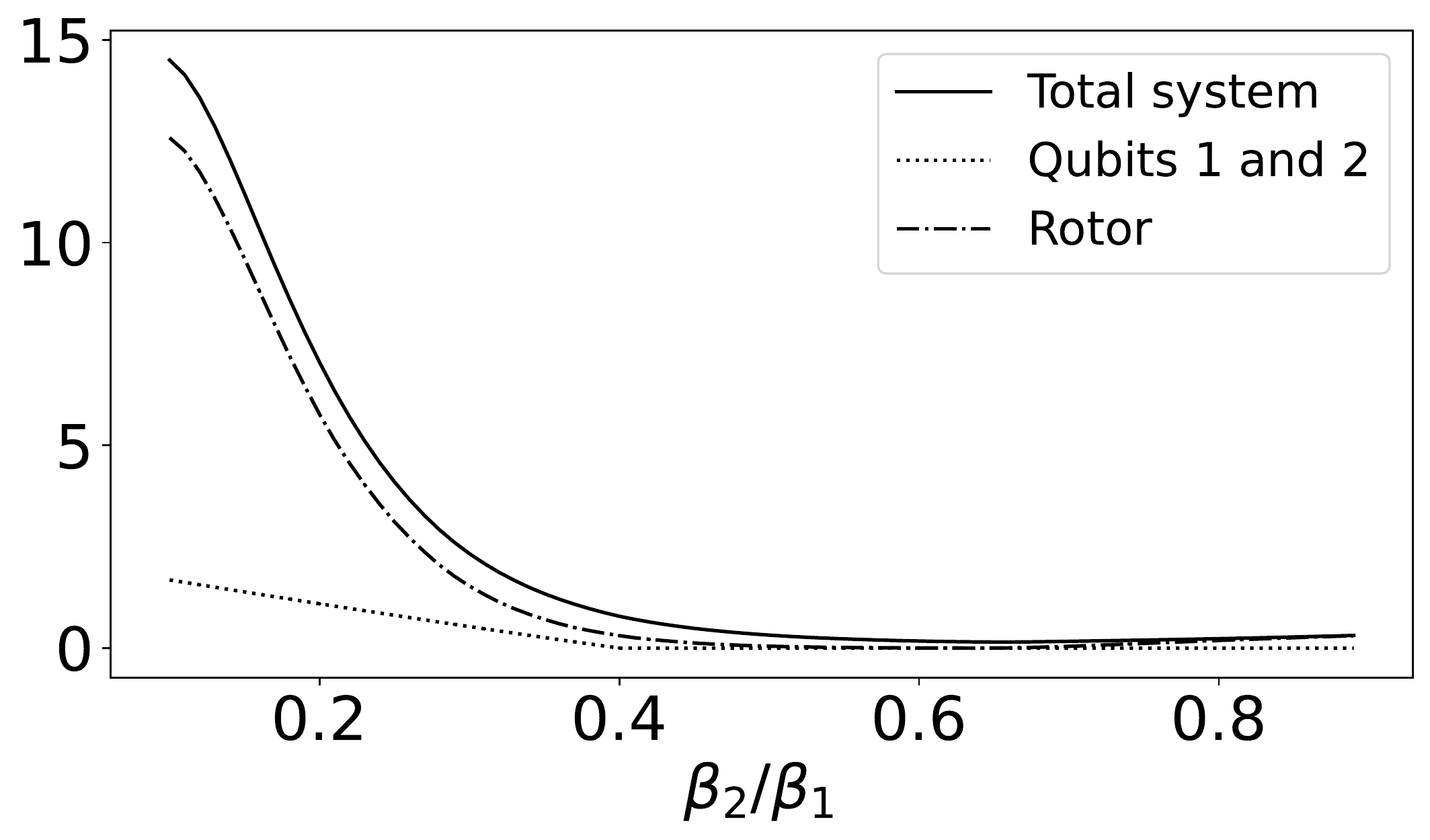}
    \caption{Ergotropy as defined in Eq. \eqref{eq:ergotropy} as a function of $\beta_2/\beta_1$. The parameters are set as in Fig. \ref{fig:tri_operations}.}
    \label{fig:3-way-ergotropy}
\end{figure}

The results provided above can be complemented by examining  the ergotropy (see Eq. \eqref{eq:ergotropy}) as a quantifier of the maximum extractable amount of work stored in the system. In Fig. \ref{fig:3-way-ergotropy}, we analyse the steady-state ergotropy
of the total system along with both qubits and the rotor, respectively, against $\beta_2/\beta_1$. The steady- state ergotropies of the two qubits separately are both zero for all $\beta_2/\beta_1$ and are therefore omitted from the plot. This is because each qubit is locally in thermal equilibrium with the bath it is coupled to and is hence in a passive state. It can be seen that for the total system, the rotor and the combined system containing the two qubits, the maximum ergotropy occurs when $\beta_2/\beta_1$ is small, corresponding to a high-temperature difference. Note that the ergotropy of the two qubits approaches zero for $\beta_2/\beta_1 \simeq B_1/B_2 = 0.4$. At this point, the populations of the second and third most populated energy levels of the two-qubit system switch and the system stays in an almost passive state as $\beta_2/\beta_1$ increases further; therefore it is unlikely that there is any significant work being extracted under unitary transformations as thermalisation occurs.

\subsection{Spin-boson model: Rectification}\label{sec:heat_rec_results}

\begin{figure}[t]
\begin{centering}
\includegraphics[width=0.95\columnwidth]{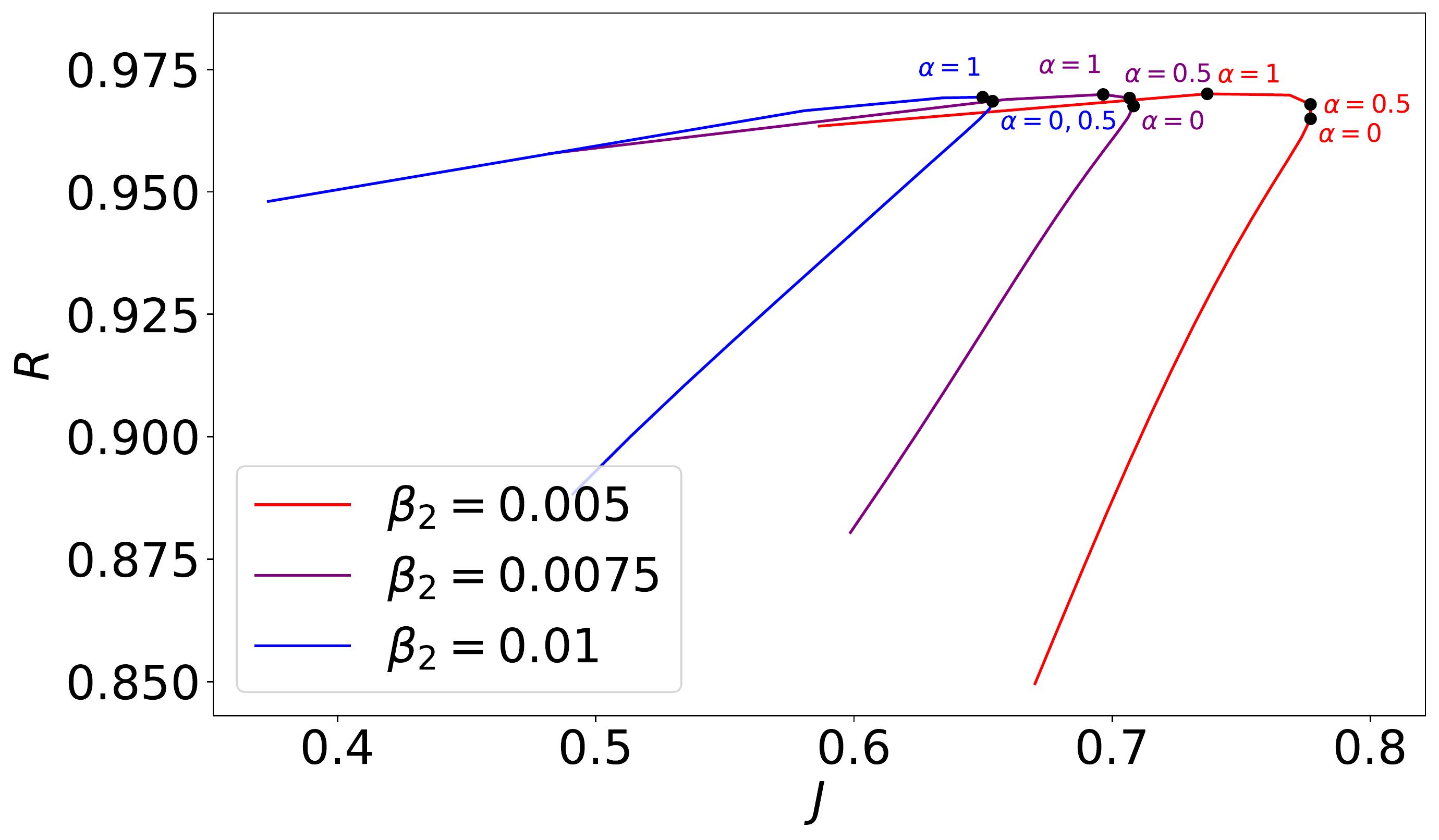}
\caption{Parametric plot of $R$ and $J$ (defined in Eqs. \eqref{eq:heatR} and \eqref{eq:heatJ}) for $0\le \chi \le 1$ at three different temperatures. The dots correspond to the values of $R$ and $J$ associated with the maximum rectification performance's measure (see Eq. \eqref{eq:rec_cop}) for a given $\alpha$. 
The parameters are: $\beta_1=0.1, \; \beta_r=0.09, \;B_1=4, \; B_2=10, \; g=0.5, \; \lambda=0.1$ and $\gamma=5\times10^{-5}$.}
\label{fig:heat_rec}
\end{centering}
\end{figure}

So far our analysis for the operation of the machine has been conducted with the use of the collision model whereas for the study of the system's rectification we instead use the spin-boson one (see Sec.~\ref{sec:GME}) with the heat flows given by Eq.~\eqref{heatflowglobal}. Under these assumptions there is no work contribution $\dot W^{(Q)}$. Figure~\ref{fig:heat_rec} displays the rectification parameter $R$ and the heat currents' magnitude $J$ as we vary $0 < \chi < 1$ for different temperatures of the hot bath $\beta_2$ and for given values of $\alpha$.
For each value of $\chi$, we calculate the corresponding heat currents into the system, before and after swapping the temperatures $\beta_1$ and $\beta_2$. We then plot the corresponding $R$ and $J$ (see Eqs. \eqref{eq:heatR} and \eqref{eq:heatJ} respectively).
As $\chi$ increases, initially so do $R$ and $J$, until a maximum $J$ is reached. Then $J$ starts to decrease correspondingly, followed by $R$. 

The dots in Fig. \ref{fig:heat_rec} show where the rectification performance's measure, $\Gamma_{\alpha}$ (see Eq. \eqref{eq:rec_cop}), is maximum for a given value of $\alpha$. If we are purely interested in having the largest heat flow into the system then we can set $\alpha=0$ and maximise $\Gamma_0 = J$, or if we focus only on the maximum rectification and not on the heat flow, we consider $\Gamma_1 = R$. Alternatively, if we value the heat flow's magnitude and rectification equally, we may want to set $\alpha =0.5$, then maximising their average value $\Gamma_{0.5}= (R+J)/2$.

From Fig. \ref{fig:heat_rec} we see that a higher temperature leads to a much larger maximum $\Gamma_0$, however, the maximum of $\Gamma_1$ is only slightly larger. Consequently, a large rectification parameter $R$ can be achieved even for a hot bath at a lower temperature at the expense of a smaller heat current.

\begin{figure}[t]
\begin{centering}
\includegraphics[width=1\columnwidth]{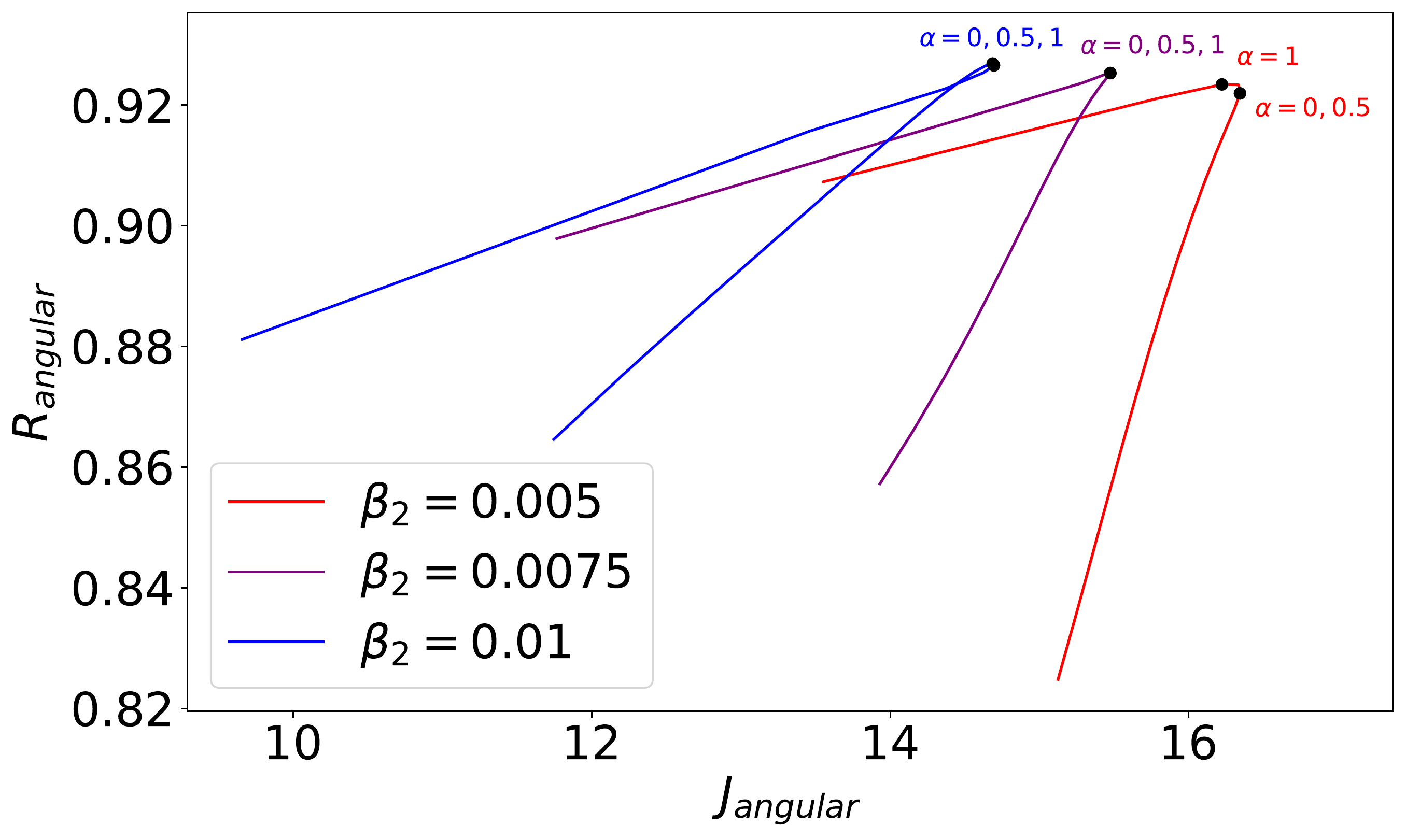}
\caption{Parametric plot of $R_{\rm angular}$ and $J_{\rm angular}$ (defined in Eqs. \eqref{eq:angmomR} and \eqref{eq:angmomJ}) for $0\le \chi \le 1$ at three different temperatures. The dots correspond to the values of $R_{\rm angular}$ and $J_{\rm angular}$ with the maximum rectification performance's measure (see Eq. \eqref{eq:rec_cop}) for a given $\alpha$.
The parameters are as in Fig. \ref{fig:heat_rec}.}
\label{fig:angmomrectri}
\end{centering}
\end{figure}

In a similar manner, we investigate the angular momentum rectification quantifiers, $R_{\rm angular}$ and $J_{\rm angular}$ using Eqs.~\eqref{eq:angmomR} and \eqref{eq:angmomJ}. Figure~\ref{fig:angmomrectri} depicts the corresponding $R_{\rm angular}$ and $J_{\rm angular}$ as $0 < \chi < 1$. We find a very similar result as in Fig. \ref{fig:heat_rec}: a large angular momentum can be achieved when bath 2 is at high temperatures. However, unlike in the case of heat rectification, we find that decreasing the temperature of the hot bath can lead to a slightly larger rectification parameter $R$.

\section{Conclusions}\label{sec:conclusions}
We studied the out-of-equilibrium  thermodynamics of a system consisting of two qubits and a quantum rotor coupled to two thermal heat baths. Two microscopic environment models were employed to analyse the dynamics, the system's steady-state and find the corresponding heat flows and external power. Various definitions of work within the rotor were discussed, and we demonstrated that the addition of a dissipative load not only provides us a way of extracting work from the rotor, but also enables the system to reach a steady state. 

Studying the steady state energy flows of the system leads to the identification of three operational regimes of the machine: an engine, a refrigerator and an accelerator. 

Finally, we investigated both the heat and angular momentum rectification by introducing a figure of merit that depends on both the rectification parameter and the maximum heat current or angular momentum. If we are more interested in the magnitude of heat flow or angular momentum then we find that a higher-temperature hot bath allows for a larger $J$ and $J_{\rm angular}$. On the other hand, for the case of heat rectification, a larger rectification parameter can be achievable with a lower temperature of the hot bath at the expense of heat, while for angular momentum rectification, we find that lowering the temperature of the hot bath can lead to a larger rectification parameter, albeit with a smaller angular momentum.

\acknowledgements
 We acknowledge  support from the UK EPSRC through Grants No. EP/S02994X/1 and 2278075. GDC and KH acknowledges support from the Royal Society through Grant No. IEC\textbackslash R2\textbackslash 222003.

\appendix

\section{Derivation of the global master equation}\label{app:globalME}
To derive the global master equation we follow standard textbooks on open quantum systems (see for instance \cite{breuer2002theory}). First, we define the eigenvalues and eigenvectors of $H_{S}$ as
\begin{equation}
H_{S}\ket{\epsilon} = \epsilon \ket{\epsilon}
\end{equation}
so that the corresponding projection operators are
\begin{equation}
\Pi (\epsilon) = \ketbra{\epsilon}{\epsilon}.
\end{equation}
In the spin-boson model, we assume each bath $i=1,2$ to consist of an ensemble of non interacting harmonic oscillators with frequencies $\omega_{i,k}$:
\begin{equation}
H_B=\sum_i\sum_k \omega_{i,k} a_{i,k}^\dagger a_{i,k}.
\end{equation}

The system-bath interaction Hamiltonian assumes the general form
\begin{equation}
H_I = \sum_{i} A_{i} \otimes E_{i}^{\dagger} + A_{i}^{\dagger} \otimes E_{i},  
\end{equation}
where $A_{i}$ and $E_{i}$ are the system's and bath's operators, respectively. We set $A_i = \sigma_i^- $ and $E_i = \sum_k g_{i,k} a_{i,k}$, $i=1,2$.
The system's  eigenoperators are defined as
\begin{equation}
A_{i}(\omega) = \sum_{\epsilon' - \epsilon = \omega} \Pi(\epsilon) A_{i}\Pi(\epsilon'),  
\end{equation}
and fulfill the relations
\begin{eqnarray}
\comm{H_{S}}{A_{i}(\omega)} &=& -\omega A_{i}(\omega),
\\
\comm{H_{S}}{A_{i}^{\dagger}(\omega)} &=& \omega A_{i}^{\dagger}(\omega).
\end{eqnarray}
In the interaction picture, we have for the system
\begin{equation}
e^{iH_{S}t}A_{i}(\omega)e^{-iH_{S}t} = e^{-i\omega t}A_{i}(\omega),
\end{equation}
and for the bath:
\begin{equation}
E_{i}(t) = e^{iH_Bt}E_{i}e^{-iH_Bt} = \sum_k g_{i,k} a_{i,k} e^{-i \omega_{i,k} t}.
\end{equation}
The two-time correlation functions are
\begin{equation}
\langle E_{i}^{\dagger}(s) E_{i}(0) \rangle = \sum_k g_{i,k}^2 e^{i \omega_{i,k} s} n_i(\omega_{i,k}),
\end{equation}
\begin{equation}
\langle E_{i}(s) E_{i}^{\dagger}(0) \rangle = \sum_k g_{i,k}^2 e^{- i \omega_{i,k} s} \left[ n_{i}(\omega_{i,k}) +1\right],
\end{equation}
where $n_{i}$ is the thermal occupation number of bath $i$ 
\begin{equation}
n_i (\omega) = \text{Tr}[a_i^{\dagger} a_i \rho_{B_i}]=\frac{1}{e^{\beta_{i} \omega }-1}.
\end{equation}
Then, an integral over a continuum of frequencies is applied
\begin{equation}
\int_0^\infty ds e^{i\omega s} \langle E_{i}^{\dagger}(s) E_{i}(0) \rangle = J_{i}(\omega) n_{i}(\omega),
\end{equation}
\begin{equation}
\int_0^\infty ds e^{i\omega s} \langle E_{i}(s) E_{i}^{\dagger}(0) \rangle = J_{i}(\omega) \left[ n_{i}(\omega) +1\right],   
\end{equation}
where $J_{i}(\omega)$ is the spectral density. Using these definitions, we can write the global master equation as 
\begin{eqnarray}\label{ME}
 \dot{\rho}_S &=& - i\comm{H_S}{\rho_S} \nonumber\\ &+& \sum_{i}\sum_{\omega} J_{i}(\omega) n_{i}(\omega) \Bigl[ A_{i}^{\dagger}(\omega) \rho_SA_{i}(\omega) \nonumber\\ 
 & &  -  \frac{1}{2} \acomm{A_{i}(\omega)A_{i}^{\dagger}(\omega)}{\rho_S}\Bigr] \nonumber\\
&+& \sum_{i}\sum_{\omega} J_{i}(\omega) \left[n_{i}(\omega) + 1\right]\Bigl[A_{i}(\omega) \rho_SA_{i}^{\dagger}(\omega)\nonumber\\ 
& & - \frac{1}{2} \acomm{A_{i}^{\dagger}(\omega)A_{i}(\omega)}{\rho_S}\Bigr].
\end{eqnarray}
We can add in the dissipative load ``locally", as in Eq.~\eqref{eq:dissload}, as long as $\gamma$ is small compared to the qubits-rotor coupling $\lambda$. This leads to the compact form of the master equation reported in Eq.~\eqref{eq:globalme}.
From here on we will choose the spectral density to be Ohmic with a cutoff frequency $\Omega$:
\begin{equation}
J_{1}(\omega) = g^2(1-\chi)^2\frac{\omega \Omega^2}{\omega^2+\Omega^2},
\end{equation}
\begin{equation}
J_{2}(\omega) = g^2(1+\chi)^2\frac{\omega \Omega^2}{\omega^2+\Omega^2},
\end{equation}
where we select $\Omega = \omega_{\text{max}}$ and $-1 \leq \chi \leq 1$.

At this stage it is worth mentioning the intense debate on the comparison between LME and GME, see for instance Refs.~\cite{Gonzalez2017,HoferNJP2017}. While alternative solutions that interpolate between the two approaches have been proposed~\cite{CattaneoNJP2019}, our GME approach coincides with the LME in the limit $\lambda\to 0$.

\begin{figure}[t]
\begin{centering}
\includegraphics[width=0.9\columnwidth]{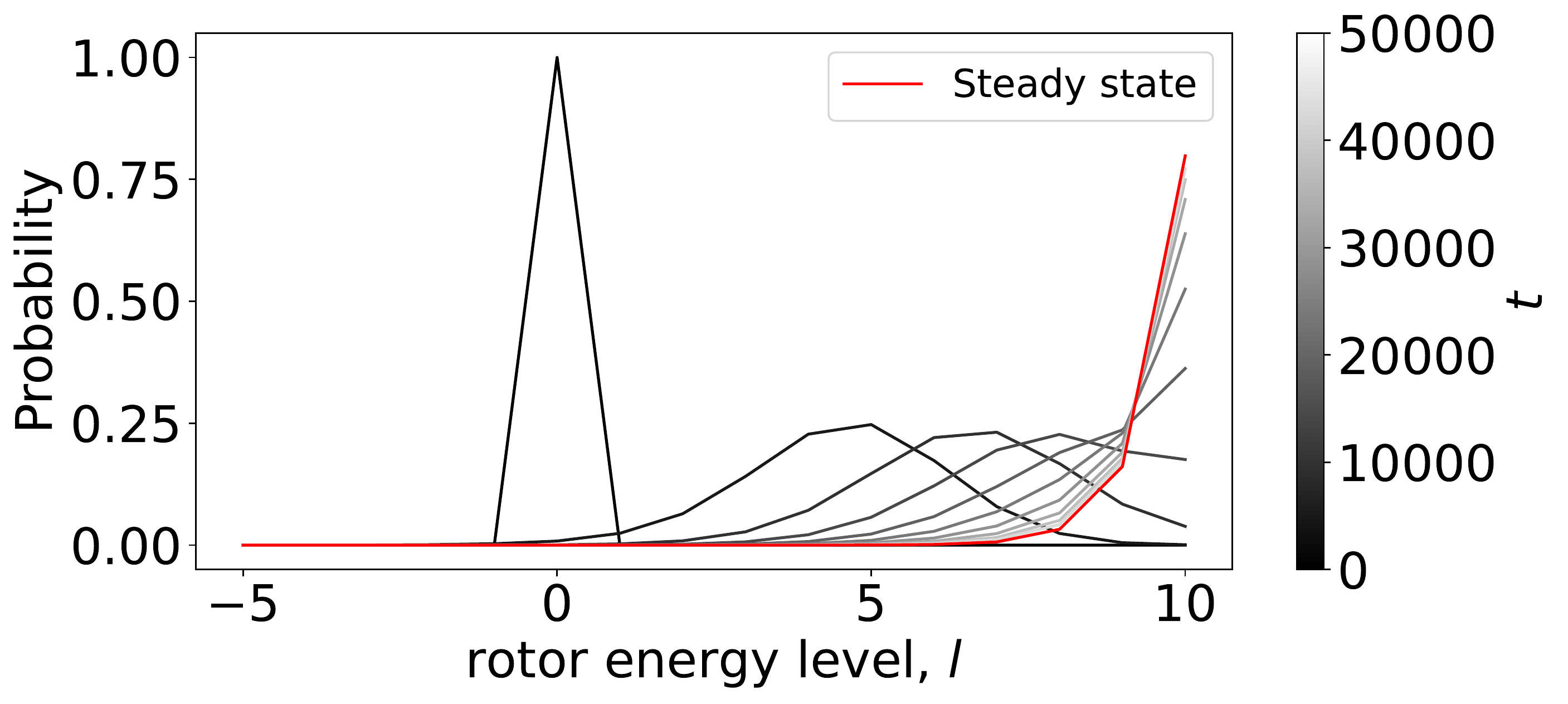}
\caption{Plots of $\bra{l} \rho_r(t) \ket{l}$ for each $-5 \leq l \leq 10$ at increasing time increments. The parameters are: $B_1 = 10, \; B_2=10, \; \beta_1 = 0.1, \; \beta_2 = 0.02, \; I = 1, \; g=1, \; \chi=0, \; \lambda = 0.1$.}
\label{fig:energylevelsnodiss}
\end{centering}
\end{figure}

\begin{figure}[t]
\begin{centering}
\includegraphics[width=0.9\columnwidth]{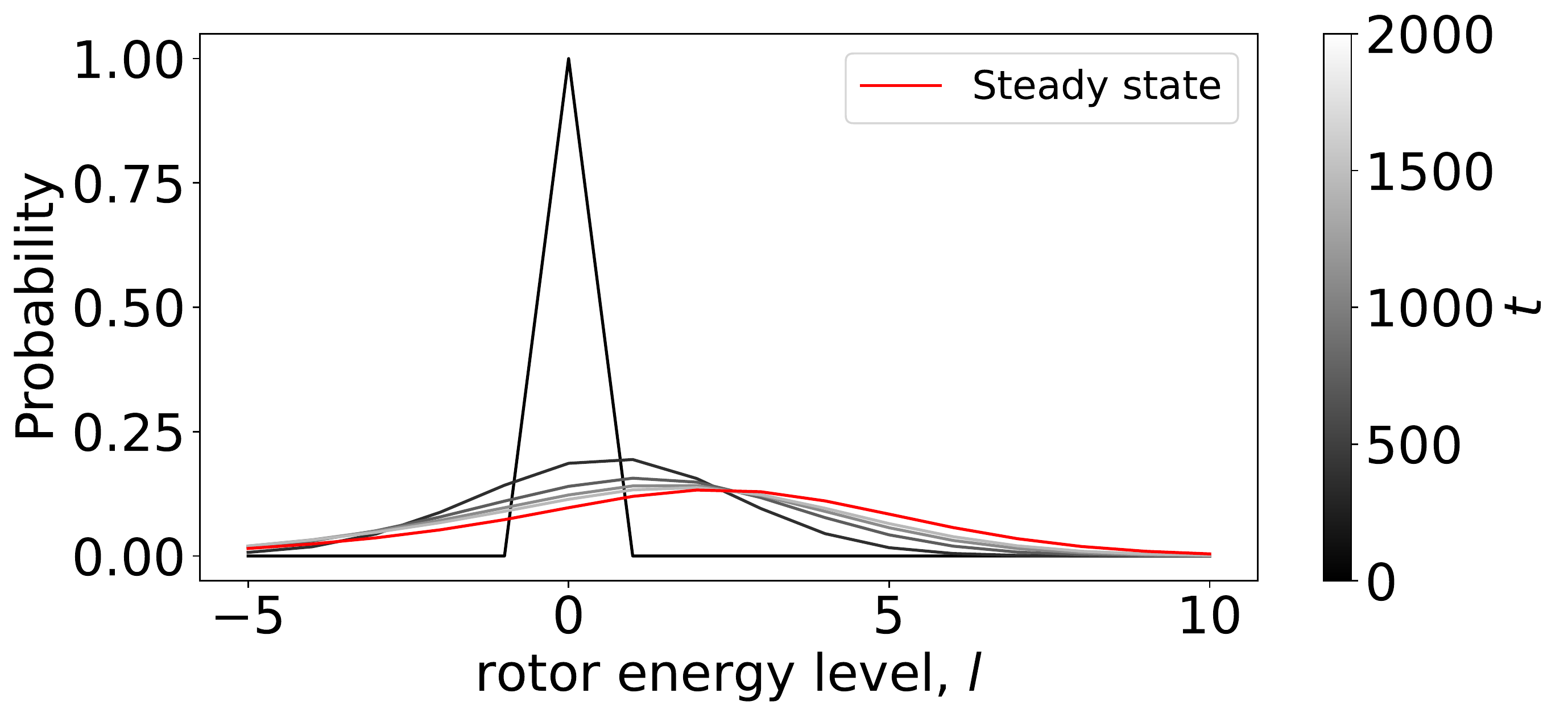}
\caption{Plots of $\bra{l} \rho_r(t) \ket{l}$ for each $-5 \leq l \leq 10$ at increasing time increments with added dissipative load. The parameters are the same as in Fig.~\ref{fig:energylevelsnodiss}, with the addition of $\gamma = 5\times10^{-4}$ and $\beta_r = 0.09$.}
\label{fig:energylevelsdiss}
\end{centering}
\end{figure}

\section{Truncation of the rotor's Hilbert space}
\label{app:truncation}
 The acceleration of the rotor can cause problems when solving the master equation numerically, as we need to truncate the Hilbert space of the rotor. We assume this to be spanned by the eigenstates $\ket l$ of the angular momentum operator $L_z$ such that $L_z\ket l = l\ket l$.
 Figure \ref{fig:energylevelsnodiss} shows how the state of the rotor evolves in time without the action of the dissipative load, modelled by the Lindblad superoperator $\mathcal L_r$. We plot the populations $\bra{l} \rho_r(t) \ket{l}$ where $\rho_r(t)$ is the reduced density matrix of the rotor obtained using the local master equation for  $-5 \leq l \leq 10$ at a given time, $t$, with the steady state shown in red. The results in Fig.~\ref{fig:energylevelsnodiss} show that the energy level most likely occupied by the rotor will increase beyond the truncated Hilbert space.

 In order to solve this problem and extract work (continuously), we add a dissipative load.
Figure \ref{fig:energylevelsdiss} shows that, with the addition of the dissipative load, the most populated energy level of the rotor does not go beyond those included in the truncated Hilbert space.

\begin{figure}[b]
    \centering
    \includegraphics[width=0.95\columnwidth]{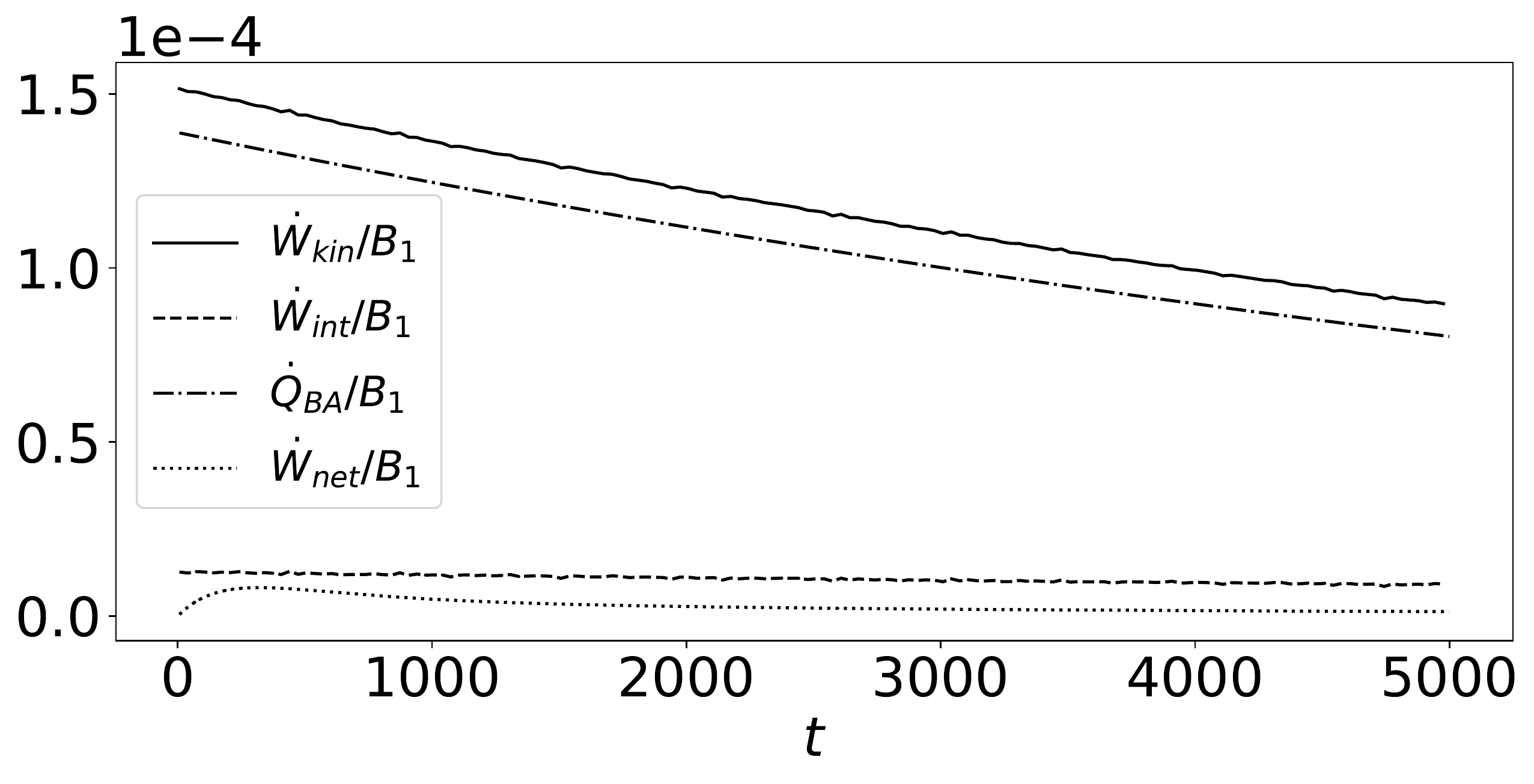}
    \caption{Evolution of the kinetic, intrinsic, net power and back action noise of the rotor in time for a system with coupling described by Eq. \eqref{eq3wayinteraction}. The steady state solutions of the net and kinetic power will be zero. The parameters are as in Fig.~\ref{fig:tri_operations} with $\beta_2/\beta_1 = 0.8$.}
    \label{fig:power-3-way}
\end{figure}

\section{Kinetic, Intrinsic and net power}\label{sec:powerappendix}

The kinetic, intrinsic and net power of the rotor (see Eqs.~\eqref{eq:kineticpower},\eqref{eq intrinsic power} and \eqref{eq net power}) are shown in Fig.~\ref{fig:power-3-way} as a function of time for a given value of $\beta_2/\beta_1=0.8$. We see that the net power quickly reaches a maximum before decreasing again, whereas the kinetic and intrinsic power steadily decrease over time. Note that the kinetic and net power will tend to zero as the system reaches its steady state, while the intrinsic power will tend to $-\dot{Q}_r$.

\bibliography{bib}
\end{document}